\documentclass[sigconf]{acmart}
\renewcommand\footnotetextcopyrightpermission[1]{} % removes footnote with conference information in first column
\usepackage{xspace}
\usepackage{fancyvrb}
\usepackage{lipsum}
\newcommand{\verbatimfont}[1]{\renewcommand{\verbatim@font}{\ttfamily#1}}
\usepackage{titlesec}

\titleformat{\section}[runin]
  {\normalfont\Large\bfseries}{\thesection}{1em}{}
\titleformat{\subsection}[runin]
  {\normalfont\large\bfseries}{\thesubsection}{1em}{}

\newcommand{\etal}{\hbox{\emph{et al.}}\xspace}

\begin{document}
\title{Automated Fix Detection Given Flaky Tests}

\author{David Landsberg}
\affiliation{\institution{University College London}}
\email{d.landsberg@ucl.ac.uk}

\author{Earl T. Barr}
\affiliation{\institution{University College London}}
\email{e.barr@ucl.ac.uk}

\maketitle

\section{Introduction}

\iffalse % latex notation-CLEAN VERSION OF INTRO for FB page upload
Developers ignore tools that they think waste their time --- hampering the adoption of verification and validation (V&V) tools in general. Automatic V&V will not be ubiquitous until we can measure its value, by answering "How many of the bugs it reports do developers fix?" Here, the problem is determining whether a fix has actually occurred given test reports --- the automated fix detection problem. Any solution is expected to be a function of a failure's symptoms, such as stack traces and user/test reports.  At Facebook, which develops software using continuous integration and deployment in conjunction with automatic V&V, the need to solve this "largely overlooked" problem is especially acute.
Alshahwan et al decompose it into two subproblems: failure grouping problem, which associates groups of failures to the methods which generate them, and proving a negative, which determines when we can be confident failures won't recur (i.e. a fix has succeeded).

We take up the challenge as follows: To group failures, we use methods of causal inference to assign each failure a root cause.  To prove a negative, we apply statistical change point detection methods to detect when a fix has succeeded in the presence of flaky tests. Combined, these offer a novel solution to the fix detection problem which is at once scalable and integratable into Facebook's development process.
\fi

%software using continuous integration and deployment~\cite{7883285,canaryrelease}.

Developers ignore tools that they think waste their time --- hampering the adoption of verification and validation (V\&V) tools in general. Automatic V\&V  will not be ubiquitous until we can measure its value, by answering "How many of the bugs it reports do developers fix?" Here, the problem is determining whether a fix has actually occurred --- the \textit{automated fix detection problem} (FDP). Any solution is expected to be a function of a failure's symptoms, such as stack traces and user/test reports.  At Facebook, which develops software using continual integration and deployment in conjunction with automatic V\&V, the need to solve this "largely overlooked" problem is especially acute~\cite{Facebook1}.
Alshahwan \etal decompose FDP into two subproblems: \emph{failure grouping}, which associates groups of failures to the methods which generate them, and \emph{proving a negative}, which determines when we can be confident failures will not recur (i.e. a fix has succeeded).

We take up this challenge: To group failures, we use methods of causal inference to assign each failure a root cause (\autoref{sec:failure:grouping}).  To prove a negative, we apply statistical change point detection methods to detect when a fix has succeeded in the presence of flaky tests (\autoref{sec:proving:neg}). Combined, these offer a novel solution to the fix detection problem which is at once scalable and integratable into Facebook's development process (\autoref{sec:deployment}).

\section{Grouping Failures}
\label{sec:failure:grouping}

The failure grouping problem (FGP) is that of grouping failures to their likely causes (here assumed to be methods). Being able to tell which failures a method causes is key to being able to tell whether it is fixed.
 Thus far, Alshahwan \etal use method identifiers (located at the top of stack traces) as the heuristic for grouping.
However, they propose this solution would be improved upon by applying techniques of causal inference.
They write "there has been much recent progress on causal inference \cite{Pearl} ... Therefore, the opportunity seems ripe for the further
development and exploitation of causal analysis as one technique for informing and understanding fix detection"~\cite{Facebook1}.

We take up Alshahwan \etal's challenge. We begin our development with 
the probabilistic measure of
causality due to Pearl~\cite{pearl2009,pearl2016causal}. We pick this particular theory because (as we shall see) there are simple and low-cost ways to estimate the value of the formula, and it opens the window to a number of different (potentially better) theories of causality.
Here, $C$ is a cause of the event $E$ when the following obtains:

\vspace*{-4.0mm}

\begin{equation}
    Pr(E | \mathit{do}(C)) > Pr(E | \mathit{do}(\neg C))
    \label{eq:pearl}
\end{equation}

The intuition is that causes raise the probability of their effects. 
Applied to FGP, we parse \autoref{eq:pearl} as follows:  $Pr(X|Y)$ reads "the
probability of $X$ given $Y$", $E$ is an event of a failure, and $C$ is the introduction of a given patch into the given
codebase. The operation $\mathit{do}(C)$ 
represents an external intervention that compels $C$ to obtain, whilst holding certain background factors fixed (in our case this is the rest of the codebase --- see Pearl for technical details~\cite{pearl2009}). Intuitively
then, $Pr(E | do(C))$ measures the probability that a
failure occurs upon the introduction of a given patch. Accordingly,
\autoref{eq:pearl} says that a
patch is a cause of the failure if the likelihood of the failure would have decreased had the patch not been introduced into the program.

A major question for our research is to estimate $Pr(E | do(C))$ and $ Pr(E |
do(\neg C))$. As a starting point, we envisage conducting a controlled experiment. Here, we assume i) we have a program together with its updated version, ii) that the updated version only differs from the original by a patch $C$, iii) that there is only one bug in the codebase, and iv) a fix for the bug repairs the method, and v) there is a test available  which can be run on both versions a given number of times  (in real-world testing scenarios we will not have to make all of these assumptions --- see~\autoref{sec:deployment}). Here, we propose $Pr(E|do(C))$  is estimated by the proportion of times the test results in failure in the updated version, and $Pr(E|do(\neg C))$  as the
proportion of times the test results in failure in the non-updated version. Note that the estimated probabilities might assume values anywhere in the interval $[0,1]$ --- depending on the presence of noise, indeterminism, flaky tests, and degree of unspecified behaviour. Accordingly, if \autoref{eq:pearl} holds, we say the method causes the given failure in that update for that test, thereby grouping the failure to the associated method as its cause. 

Pearl's theory is not enough. 
It is not guaranteed to handle (what Alshahawan calls) \textit{false grouping}~\cite{Facebook1}.
Accordingly, \autoref{eq:pearl} may include too many (or too few) causes in practice. To investigate this, we propose 
experimenting with different measures for the \textit{degree of causality} (which in our context may be said to measure \textit{error-causing degree}), such as $Pr(E | do(C)) - Pr(E | do(\neg C))$ and $Pr(E | do(C))/Pr(E | do(\neg C))$~\cite{pearl2009}, and saying causality obtains when the value given by the measure is over a given bound.  
Previous research has confirmed that different measures of causality perform very differently~\cite{eval}, suggesting a requirement to experiment with many different measures from the literature on A.I., fault localisation, and philosophy of science, of which there are hundreds~\cite{eval}.  

\vspace*{-5.0mm}

\section{Proving a Negative}
\label{sec:proving:neg}

Alshahwan \etal ask the following: "how long  should we wait, while continually observing no re-occurrence of a failure (in testing or production)
before we claim that the root cause(s) have been fixed?"~\cite{Facebook1}  Here, we assume the root cause(s) of a failure have been estimated by the work of~\autoref{sec:failure:grouping}. The famous \textit{proving a negative problem} rears its head here: How we can prove a negative (no more failures) in the absence of direct evidence to the contrary.  
Alshahwan \etal state that identifying the correct \textit{fix detection protocol}~\cite{Facebook1} provides the solution, and experiment with their own protocol within the Sapienz Team
at Facebook. Their protocol uses heuristics and a finite state machine, but emphasize they "do not claim it is the only possible protocol, nor that it is best among alternatives".  Accordingly, In this section we propose an alternative.

We begin our development by answering Alshawan's question above directly: We wait until we can claim a fix has occurred, i.e. when the error-causing behaviour of the method has diminished. 
Our answer is made precise as follows. 
We let the \textit{error causing behaviour} of a given method be a time series T = $t_1, t_2, \dots, t_n$, where each datapoint is an error-causing degree for a given failure group (as per \autoref{sec:failure:grouping}) over a given period. Let T1 = $t_1, t_2, \dots , t_k$ and T2 = $t_{k+1}, t_{k+2}, \dots , t_n$ be two adjacent time series splitting T.  Following the standard definition of changepoint detection, a \textit{changepoint} is detected for T1 and T2 if T1 and T2 are shown to be drawn from a different distribution according to a given hypothesis testing method~\citep{Aminikhanghahi, arXiv:1411.7955}.
We detect that some \textit{fix}/\textit{bug} has been introduced into T2 since T1, if i) a changepoint is detected for T1 and T2 and ii) the average error causing degree in T2 is smaller/larger than the average error causing degree in T1. Finally, we say the the error-causing behaviour of the method has \textit{diminished} when a fix is detected.

\begin{figure}[t!]
  \includegraphics[width=\linewidth]{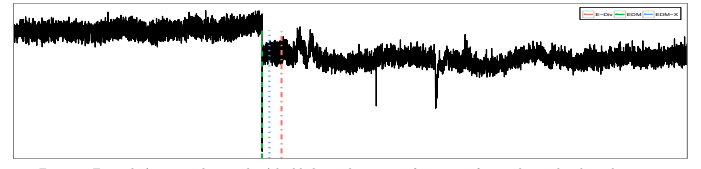}
  \caption{Time Series with Change Point.}
  \label{f1}
  
\vspace*{-0.6cm}

\end{figure}

To illustrate the setup, consider~\autoref{f1}, which represents a time series of real-valued datapoints. Let T1 be the series before the vertical green line and T2 the series after. Already, our setup could be used to say some fix has been introduced into T2 since T1. It then remains to find the precise point where the fix was introduced. This is done by applying a changepoint detection method (CDM). In general, CDMs try to identify exact times (\textit{changepoints}) when the probability distribution of a stochastic process or time series can be confidently said to change. 
Ideally, we would apply a CDM which identifies the changepoint with the datapoint indicated by the green line in~\autoref{f1}.  
Research into CDMs is a large and well-developed area~\citep{Aminikhanghahi, arXiv:1411.7955}, and have been applied successfully to solve similar problems to FDP in continuous code deployment~\cite{arXiv:1411.7955}. Key differences between CDMs include where they locate changepoints, and how scalable the technique is.

\vspace*{-0.2cm}

\section{Deployment}
\label{sec:deployment}

We first discuss three integration scenarios; with the Sapienz tool, FBlearner, and canary testing. We then discuss the development of our techniques.

The first area of deployment is alongside the Sapienz tool, which
has been integrated alongside Facebook's production development process Phabricator~\citep{Facebook1} to help identify faults. Accordingly, our methods could be integrated alongside Sapienz to help detect fixes made as a consequence of testing. 
The second area of deployment is alongside FBLearner, a Machine Learning (ML) platform through which most of Facebook's ML work is conducted. In FBlearner there is an existing fix detection workflow stage~\citep{Facebook1}, which involves using reinforcement learning to learn to classify faults and fixes. Accordingly, our methods could be integrated in the fix classification stage. 
The third area of deployment is alongside Facebook's canary testing/rolling deployment process for mobile devices. Canary releasing slowly rolls out changes to a small subset of users before rolling it out to the entire infrastructure. Facebook uses a strategy with multiple canaries (versions)~\cite{7883285,canaryrelease}.
In practice, data about different canaries could be used to form part of the dataset used for our fix detection methods. Namely, if an update is deployed in one cluster but not another, we will have important data about which failures are caused by which updates and for which methods. 

We now discuss development issues. To develop 2.1, we will need an experimental framework where we can evaluate the performance of different causal measures on given benchmarks using standard IR measures (such as accuracy, precision, recall, and F-scores). We will evaluate the measures on different testing scenarios which do not make many of the restrictive assumptions outlined in 2.1. For instance, if i) is not true we need to perform fault localisation using a causal measure on the updated program alone (using a given fault localisation setup~\cite{eval}). If ii) or iii) are not true we will need to employ measures empirically demonstrated to perform well in the presence of noise~\citep{pearl2016causal}. 

The development of 2.2 will include an experimental comparison of different CDMs, testing for effectiveness and scalability when employed at the fix detection task. To measure effectiveness, we use standard IR methods~\citep{Aminikhanghahi, arXiv:1411.7955}. To measure scalability, we will measure practical runtime on representative benchmarks. This work is made feasible insofar as many CDMs are already implemented, known to scale well, and can be used in an "online" contexts involving continuous real-time streams of datapoints.\footnote{\url{http://members.cbio.mines-paristech.fr/~thocking/change-tutorial/RK-CptWorkshop.html}}

\bibliographystyle{ACM-Reference-Format}
\bibliography{sample-bibliography}

\pagebreak

\end{document}